\documentclass{article}

     \PassOptionsToPackage{numbers, compress}{natbib}

\usepackage[numbers]{natbib}

\usepackage[final]{neurips_2024}

\usepackage[utf8]{inputenc} 
\usepackage[T1]{fontenc}    
\usepackage{hyperref}       
\usepackage{url}            
\usepackage{booktabs}       
\usepackage{amsfonts}       
\usepackage{nicefrac}       
\usepackage{microtype}      
\usepackage{xcolor}         

\title{Provocation on Expertise in Social Impact Evaluations of Generative AI (and Beyond)}

\author{
  Zoe Kahn\\
  School of Information \\
  UC Berkeley \\
  \texttt{zkahn@berkeley.edu} \\
   \And
   Nitin Kohli \\
   Center for Effective Global Action \\
   UC Berkeley  \\
   \texttt{nitin.kohli@berkeley.edu} \\
}

\begin{document}

\maketitle

\begin{abstract}
Social impact evaluations are emerging as a useful tool to understand, document, and evaluate the societal impacts of generative AI. In this provocation, we begin to think carefully about the types of experts and expertise that are needed to conduct robust social impact evaluations of generative AI. We suggest that doing so will require thoughtfully eliciting and integrating insights from a range of \textit{domain experts} and \textit{experiential experts}, and close with five open questions. 
\end{abstract}

\section{Introduction}
Social impact evaluations (e.g., \citep{solaiman2023evaluating}) are emerging as a useful tool to understand, document, and evaluate the societal impacts of generative AI; these evaluations take an important step towards the responsible development and enhanced accountability of artificial intelligence. Who develops categories of social impact---and who conducts these evaluations---shapes how generative AI is evaluated and held accountable. In this provocation, we begin to think carefully about the types of experts and expertise needed to conduct robust social impact evaluations. Without adequate expertise, there is a risk of developing evaluation criteria that fail to capture real-world harms or producing misleading evaluations that obscure critical issues. In this provocation we suggest that robust social impact evaluations require eliciting and integrating input from ``domain experts'' alongside ``experiential experts.'' As working definitions, domain experts refer to people who have received training or professionalization in a particular domain such as data science, legal studies, history, among others; experiential experts refer to ``people who are living the experience or those closely associated with someone living the experience'' \citep{young2019toward}. In our use of the term, experiential experts are not only experts in their own lived experience but also experts in culture \citep{Abebe_2021}, organizational structure \citep{pfeffer1977information}, local values \citep{Abokhodair2016Privacy}, among others.

\section{Airplane Design: An Illustrative Analogy}
We motivate the discussion of experts and expertise in social impact evaluations of generative AI by turning attention to the design of a physical artifact: the airplane. Let’s imagine we wanted to build an airplane that was safe to fly and comfortable for passengers. To do so, we would need to elicit and integrate insights from different types of experts: structural engineers to make sure the plane is strong enough to withstand the loads it could encounter; aerodynamicists to ensure the plane is as efficient and aerodynamic as possible; and experiential experts---or passengers---to make sure that the plane is comfortable. It would be important that each expert provide input on the aspects that align with their expertise---we would want structural engineers to weigh in on their particular area of specialization, but not define what constitutes passenger comfort; and we would want passengers to weigh in on the comfort of the plane, not the structural design of the plane.



\section{Attending to Expertise in Social Impact Evaluations of Generative AI}
While airplane design is distinct from generative AI, attending to expertise can inform how robust social impact evaluations are conducted in practice. What we find appealing about the framing of expertise is that it treats all types of experts with dignity and provides a mechanism for broad accountability: no one kind of expert is necessarily more important than another.  


\textit{Provocation 1: On the Experts Needed to Evaluate Social Impact.} Each category of social impact likely requires different types of experts. Evaluating  environmental costs may require individuals with expertise calculating water and energy consumption, a type of expertise not relevant to privacy and data protection evaluations. Even within a single social impact category (e.g., privacy and data protection), evaluations will likely require appropriate combinations of experts. While domain experts---such as technical experts to measure model memorization---are important, they may be insufficient on their own. Building on our prior work in rural Togo and a long lineage of work demonstrating that local norms and expectations of privacy depend on the social, political, and cultural context \citep{Abebe_2021, Abokhodair2016Privacy}, we find that the data privacy concerns raised by domain experts often fail to include data privacy concerns of experiential experts; in our case, rather than concerns about the use of personal data in algorithms, individuals in rural Togo raised relational privacy harms that can arise when sensitive data is revealed to people nearby such as family members or people in the community \citep{kahn2025privacy}. Here, and more broadly, robust social impact evaluations will likely hinge on including the appropriate combination of experts; failing to include the appropriate experts---or asking experts to weigh in on topics that do not align with their expertise---could obscure critical issues and lead to misleading evaluations. 

\textit{Provocation 2: On the Construction of the Social Impact Evaluation Framework.} Attending to  expertise in the construction of the social impact evaluation framework reveals key strengths and potential limitations. Consider that the social impact evaluation framework was developed by a group of nearly 60 domain experts spanning industry, academia, civil society, and government \citep{solaiman2023evaluating}. This is a strong, diverse range of expertise incorporated into its creation.  It also reveals the types of experts who were not included, such as experiential experts. Such absence likely has implications for the categories of social impact that are (and are not) part of the social impact evaluation framework. For example, experiential experts in low- and middle-income countries may expand definitions of existing categories of social impact or reveal entirely new categories. Again, building on the first-author's prior work in rural Togo, new categories may be needed to capture how generative AI can shift interpersonal relationships and community dynamics  \citep{kahn2025cohesion}. Importantly, observing which experts were part of the process positions us to ask: What experts are missing? What new categories are needed? How else might such frameworks need to shift or adapt?

\section{Five Open Questions}
In this provocation, we begin to think carefully about the types of experts and expertise needed to conduct robust social impact evaluations of generative AI; we close with five open questions. See \textit{Appendix  \ref{app:togo_background}} on our work in Togo and \textit{Appendix \ref{app:differences_with_design}} drawing comparisons to design approaches. More broadly, while the paper focuses on social impact evaluations of generative AI we anticipate much of this thinking, including the open questions, applies to the design of sociotechnical systems writ large.   

\begin{enumerate}

    \item \textbf{Conceptualizing experts.}  Beyond domain experts and experiential experts, are other categories of experts needed for a comprehensive conceptualization of expertise?  
     \item \textbf{Identifying experts and expertise.} How do you identify which experts and expertise are needed? How many experts of each type are enough? How do you identify the questions each type of expert is uniquely positioned to provide input on? Who decides? 
    \item\textbf{Enabling input.} How is meaningful input elicited from different types of experts? When is it important for experts to understand technical details? How are these technical details communicated in ways that enable meaningful input from experts with diverse expertise (particularly experts who may lack technical backgrounds)? 
    \item \textbf{Defining harms.} What counts as a harm? How are harms prioritized? Who decides? 
    \item\textbf{Resolving tensions.} How are tensions resolved when experts disagree? Tensions may arise within a type of expert (e.g., privacy experts disagree with one another) or between types of experts (e.g., privacy experts may raise concerns in tension with experiential experts).

\end{enumerate}

\begin{ack}
Thank you to Christopher Jirucha and Geoff Martindale for providing aerospace expertise to ensure the accuracy of the airplane analogy. 
\end{ack}



\bibliographystyle{ieeetr}
\bibliography{references}

\newpage
\appendix

\section*{Appendices}
\section{From Theory to Practice: Identifying, Eliciting and Integrating Expertise from Experiential and Domain Experts}
\label{app:togo_background}

To begin to move from theory to practice for how experts and expertise can be effectively identified, deployed, and integrated into evaluation frameworks, we provide an example from our own work leveraging experts and expertise to study data privacy in low-and middle-income countries \citep
{kahn2025privacy} and reflect on possible implications for social impact evaluations of generative AI. While data privacy is not the same as social impact evaluations of generative AI, the structure and process may be informative for social impact evaluations of generative AI. In particular, social impact evaluations of generative AI may similarly find it useful to: identify experiential and domain experts and expertise currently represented; identify missing experiential and domain experts and expertise; initiate intentional efforts to elicit input from missing experts; ask experts for input on topics that leverage their unique expertise; and integrate across experts and expertise. This should be seen as the start of a conversation, not an articulation of best practices. 

\textbf{Research context.} To situate our prior work, in response to the Covid-19 pandemic, the Government of Togo launched the world’s first entirely digital cash transfer program that used machine learning and mobile phone metadata to determine program eligibility \citep{aiken2022machine}. We, along with our Togolese colleague and colleagues in data science and economics who trained the machine learning models and collaborated with the Government of Togo to implement the program, wanted to understand the data privacy risks that arise from the use of mobile phone metadata as part of development interventions. To do this work, we followed the following process. Importantly, while the process appears linear in its written form below, it was an iterative process in practice. 

\begin{enumerate}
    \item\textit{Identify experiential and domain experts and expertise currently represented.} In our work in rural Togo, a literature review revealed that privacy and development experts and expertise were primarily represented in the literature exploring the data privacy concerns related to mobile phone metadata and its application to development. Leveraging what we learned in rural Togo, social impact evaluations of generative AI may benefit from identifying which types of experts and expertise are initially represented in evaluations. Methodologically, this could be done through a literature review or documenting in real-time which experts and expertise are represented in a given evaluation, among others. 
    \item\textit{Identify important missing experiential and domain experts and expertise.} In our work in rural Togo, after documenting the experts and expertise currently represented it was notable that experiential experts were missing, namely, people living in rural Togo. Given that we were interested in data privacy, it was important to have experiential experts represented because, not only were their data used, but it is well documented that privacy norms differ across cultures. Leveraging what we learned in rural Togo, social impact evaluations of generative AI may similarly benefit from identifying the delta between the types of experts and expertise needed to conduct a given evaluation and the types of experts and expertise currently represented. It may also be that the necessary types of experts and expertise change over the course of a social impact evaluation, such as if the evaluation is taking place before or after AI release, requiring periodic re-evaluation. 
    \item\textit{Initiate intentional efforts to elicit input from missing experts.} In our work in rural Togo, to understand the data privacy concerns of experiential experts the first author conducted ethnographically-informed fieldwork and semi-structured interviews in rural Togo to understand the data privacy concerns of people living in rural villages related to the collection and use of mobile phone metadata. Leveraging what we learned in rural Togo, social impact evaluations of generative AI may also benefit from intentional efforts to elicit input from missing experts. Methodologically, the most appropriate methods may depend on the type of expert (e.g., participant observation, interviews, expert panels, statistical reports, requests for feedback on system design, among others).
    \item\textit{Ask experts for input on topics that leverage their unique expertise.} In our work in Togo, we initially got it wrong: we asked questions that did not align with the expertise of our participants in rural Togo. Inspired by Future Workshops \citep{kensing2020generating}, our initial interviews asked participants to envision different uses of mobile phone data, both uses that should be supported and those that should be prevented. Our participants living in rural Togo often responded with blank stares, telling us they had little understanding of mobile phone data, let alone how it could be used or misused. This made us recognize our questions were perhaps better suited to a person with expertise envisioning data uses. After reflecting, we came to understand that experiential experts were especially well positioned to situate data within their everyday lives. We pivoted, first developing new methods that leverage visuals and storytelling to explain the data recorded in mobile phone metadata, helping participants with varying levels of literacy and formal education to develop an intuition for how that data could be used to draw inferences \citep{kahn2025privacy}; then, shifting to ask participants how they would feel if their mobile phone data were shared with their spouse, household, village, the government, or researchers. This was a topic participants had a lot to say about, in part, because it aligned with their area of expertise. Leveraging what we learned in rural Togo, social impact evaluations of generative AI will likely be more robust if experts are asked to provide input on topics that align with their expertise, and in some cases it may be important to scaffold technical details to enable meaningful input. 
    \item\textit{Integrate across experts and expertise}. In our work in rural Togo, we found that people in rural Togo surfaced a different set of privacy concerns than people with privacy and development domain expertise. In contrast to domain expert concerns related to surveillance, consent, and technical re-identification, individuals living in rural Togo raised a set of relational privacy concerns that could arise if mobile phone data were revealed to people nearby such as spouses, households, or the broader village. We do not argue that the privacy concerns of experiential experts are “right” and domain experts “wrong” (or vice versa). Instead, we take an integrative approach and demonstrate that addressing data privacy holistically will require reckoning with the data privacy concerns raised by both domain experts and experiential experts. Leveraging what we learned in rural Togo, social impact evaluations of generative AI will likely also encounter instances where different experts (of the same type or of different types) surface different concerns or provide different evaluations that are in tension with one another. In instances of disagreement, it will likely be important to identify constructive paths forward that take expertise into account and treat experts with dignity. Depending on the topic, this may entail deferring to one type of expert or expertise or integrating across different types of experts and expertise.
\end{enumerate}

In this appendix, we provided an example from our own work leveraging experts and expertise to study data privacy in rural Togo, reflecting on how the lessons learned can be leveraged to inform social impact evaluations of generative AI. Effectively attending to experts and expertise in social impact evaluations of generative AI is an exciting opportunity for creativity and innovation. More broadly, we intuit that careful attention to experts and expertise may be  applicable to sociotechnical system design writ large. 

\section{Centering Experts and Expertise: Related To (And Distinct From) Established Design Approaches}\label{app:differences_with_design}

As part of an emerging body of work examining experts and expertise in sociotechnical system design \citep{turing2022stakeholder, birhane2022power, groves2022algorithmic, impact2022assessment, zytko2022participatory, delgado2023participatory, PakNDintroducing}, we delineate how our approach builds on, but is distinct from, several well established design traditions. By attending to experts and expertise, we center \textit{who} participates in the design process, with a particular focus on \textit{what} expertise those individuals possess so that sociotechnical systems support a broad range of human values. In what follows we begin to articulate how attending to experts and expertise relates to two established design traditions, user centered design (UCD) and value sensitive design (VSD). We caution that this discussion is  preliminary and additional work is needed to properly explore experts and expertise in theory and practice. 

UCD engages users so that sociotechnical systems are more useful and usable \citep{norman1986user}. We expand UCD in two ways: 
\begin{enumerate}
    \item\textit{Expand who participates.} The focus on experts and expertise expands who participates in the design process beyond users to a broader group of experts. This includes domain experts (e.g., privacy scholars, historians, lawyers) as well as experiential experts who may be users or non-users alike (e.g., people with lived expertise navigating low-bandwidth settings, people with lived expertise choosing not to use a technology, people with lived expertise who are unable to use  a technology for any number of reasons). 
    \item\textit{Expand the purpose of participation.} Traditionally, UCD engages users so that products are more useful and usable to end users. While these are two important values, we believe assembling an appropriate set of experts can be leveraged to envision, develop, and evaluate sociotechnical systems in ways that support a broader range of human values (e.g., dignity, autonomy, privacy, accountability, among others). 
\end{enumerate}

VSD engages direct and indirect stakeholders so that sociotechnical systems support a broad range of human values \citep{friedman2019value}. While more work is needed to theorize experts and experts, we suspect centering experts and expertise is complementary to VSD. 
\begin{enumerate}
    \item\textit{Complement stakeholders with experts and expertise to inform constructive paths forward.} VSD engages direct and indirect stakeholders, identified by role, in large part to understand how sociotechnical systems may implicate human values to inform constructive paths forward. By contrast, attending to experts and expertise turns more explicit attention toward whose knowledge is needed to inform constructive paths forward. While there is some overlap in who participates based on these two approaches, the rationale for participation is somewhat different and, correspondingly, what is being asked is different. To make this concrete, let us imagine that we are conducting a social impact evaluation of generative AI related to environmental costs and carbon emissions \citep{solaiman2023evaluating}. VSD would have us conduct a stakeholder analysis to identify direct and indirect stakeholders based on their role and stake in the environmental costs and carbon emissions of generative AI. Through such a process, we might identify a range of stakeholders including environmental advocates, people living in areas especially impacted by climate change, people living in areas near data centers and other large scale compute resources, and non-human stakeholders such as rivers, oceans, and creatures from small to large. This would help us begin to understand who may be impacted and how. By contrast, a focus on experts and expertise would have us consider whose knowledge is needed to conduct such a social impact evaluation. Through such a process, we might identity people trained to calculate carbon emissions, people with lived expertise who live nearby large infrastructure and can speak to the impact on their lives, people with historical expertise who can help understand how the introduction of infrastructure shifts place over time, and people with emerging expertise representing the perspectives of non-human stakeholders in sociotechnical design processes. This example begins to illustrate how an approach that attends to stakeholders, experts, and expertise is likely better positioned to construct paths forward than any on their own. 
    \item\textit{Focus on the topics each stakeholder (or expert) is well positioned to weigh in on.} In light of the somewhat different rationales for participation, VSD does not address if there are topics that particular stakeholders are more (or less) well positioned to weigh in on. In contrast, we intuit that different experts will be well positioned to weigh in on different topics, and the successful implementation of this approach will likely hinge on engaging experts on topics that align with their expertise. 
\end{enumerate}

In this appendix, by delineating between existing design approaches (i.e., UCD and VSD), we begin to articulate how attending to experts and expertise in social impact evaluations of generative AI (and beyond) may provide new, constructive, and complementary paths forward.  
\newpage
\section*{NeurIPS Paper Checklist}

\begin{enumerate}

\item {\bf Claims}
    \item[] Question: Do the main claims made in the abstract and introduction accurately reflect the paper's contributions and scope?
    \item[] Answer: \answerYes{} 
    \item[] Justification: In the abstract, we claim to attend to experts--and expertise--in social impact evaluations of generative AI; that is the contribution of the paper. 

\item {\bf Limitations}
    \item[] Question: Does the paper discuss the limitations of the work performed by the authors?
    \item[] Answer: \answerNo{} 
    \item[] Justification: This provocation makes clear that attending to expertise in social impact evaluations is early work and therefore the limitations are not yet known. To that end, we end the paper with five open questions which point to areas for future exploration and research.

\item {\bf Theory Assumptions and Proofs}
    \item[] Question: For each theoretical result, does the paper provide the full set of assumptions and a complete (and correct) proof?
    \item[] Answer: \answerNA{} 
    \item[] Justification: This paper does not provide any theoretical results. 

    \item {\bf Experimental Result Reproducibility}
    \item[] Question: Does the paper fully disclose all the information needed to reproduce the main experimental results of the paper to the extent that it affects the main claims and/or conclusions of the paper (regardless of whether the code and data are provided or not)?
    \item[] Answer: \answerNA{} 
    \item[] Justification: This paper does not include any experiments. 

\item {\bf Open access to data and code}
    \item[] Question: Does the paper provide open access to the data and code, with sufficient instructions to faithfully reproduce the main experimental results, as described in supplemental material?
    \item[] Answer: \answerNA{} 
    \item[] Justification: This paper does not include experiments requiring code. 

\item {\bf Experimental Setting/Details}
    \item[] Question: Does the paper specify all the training and test details (e.g., data splits, hyperparameters, how they were chosen, type of optimizer, etc.) necessary to understand the results?
    \item[] Answer: \answerNA{} 
    \item[] Justification: This paper does not include any experiments. 

\item {\bf Experiment Statistical Significance}
    \item[] Question: Does the paper report error bars suitably and correctly defined or other appropriate information about the statistical significance of the experiments?
    \item[] Answer: \answerNA{} 
    \item[] Justification: This paper does not include any experiments.

\item {\bf Experiments Compute Resources}
    \item[] Question: For each experiment, does the paper provide sufficient information on the computer resources (type of compute workers, memory, time of execution) needed to reproduce the experiments?
    \item[] Answer: \answerNA{} 
    \item[] Justification: This paper does not include any experiments. 
    
\item {\bf Code Of Ethics}
    \item[] Question: Does the research conducted in the paper conform, in every respect, with the NeurIPS Code of Ethics \url{https://neurips.cc/public/EthicsGuidelines}?
    \item[] Answer: \answerYes{} 
    \item[] Justification: We have reviewed the NeurIPS Code of Ethics and confirm that the paper conforms to the code. 

\item {\bf Broader Impacts}
    \item[] Question: Does the paper discuss both potential positive societal impacts and negative societal impacts of the work performed?
    \item[] Answer: \answerNo{} 
    \item[] Justification: Given the short (2-page) length of the provocation, we do not include a separate section on broader impacts. However, the paper attends to the types of experts---and expertise---to conduct more robust social impact evaluations of generative AI, and ends with a set of questions that point to potential limitations. 
    
\item {\bf Safeguards}
    \item[] Question: Does the paper describe safeguards that have been put in place for responsible release of data or models that have a high risk for misuse (e.g., pretrained language models, image generators, or scraped datasets)?
    \item[] Answer: \answerNA{} 
    \item[] Justification: This paper poses no such risks. 

\item {\bf Licenses for existing assets}
    \item[] Question: Are the creators or original owners of assets (e.g., code, data, models), used in the paper, properly credited and are the license and terms of use explicitly mentioned and properly respected?
    \item[] Answer: \answerNA{} 
    \item[] Justification: This paper does not use existing assets. 

\item {\bf New Assets}
    \item[] Question: Are new assets introduced in the paper well documented and is the documentation provided alongside the assets?
    \item[] Answer: \answerNA{} 
    \item[] Justification: This paper does not release new assets. 

\item {\bf Crowdsourcing and Research with Human Subjects}
    \item[] Question: For crowdsourcing experiments and research with human subjects, does the paper include the full text of instructions given to participants and screenshots, if applicable, as well as details about compensation (if any)? 
    \item[] Answer: \answerNA{} 
    \item[] Justification: This paper does not involve crowdsourcing nor research with human subjects. 

\item {\bf Institutional Review Board (IRB) Approvals or Equivalent for Research with Human Subjects}
    \item[] Question: Does the paper describe potential risks incurred by study participants, whether such risks were disclosed to the subjects, and whether Institutional Review Board (IRB) approvals (or an equivalent approval/review based on the requirements of your country or institution) were obtained?
    \item[] Answer: \answerNA{} 
    \item[] Justification: This article does not involve crowdsourcing nor research with human subjects. 

\end{enumerate}


\end{document}